\documentclass[pdflatex,sn-mathphys]{sn-jnl}

\setcitestyle{super,open={},close={},citesep={,}}
\makeatletter

\usepackage{anyfontsize}
\usepackage{graphicx}%
\usepackage{multirow}%
\usepackage{amsmath,amssymb,amsfonts}%
\usepackage{amsthm}%
\usepackage{mathrsfs}%
\usepackage[title]{appendix}%
\usepackage{xcolor}%
\usepackage{textcomp}%
\usepackage{manyfoot}%
\usepackage{booktabs}%
\usepackage{algorithm}%
\usepackage{algorithmicx}%
\usepackage{algpseudocode}%
\usepackage{listings}%
\usepackage{hyperref}
\usepackage[symbol]{footmisc}
\usepackage{xcolor}%

\newcommand\lamobs{$\lambda_{\rm{obs}}$} 
\newcommand\lamrest{$\lambda_{\rm{rest}}$}

\begin{document}

\title[Article Title]{Extended hot dust emission around the earliest massive quiescent galaxy}


\author*[1]{\fnm{Zhiyuan} \sur{Ji}}\email{zhiyuanji@arizona.edu}

\author[2,1]{\fnm{Christina C.} \sur{Williams}}

\author[1]{\fnm{George H.} \sur{Rieke}}

\author[1]{\fnm{Jianwei} \sur{Lyu}}

\author[1]{\fnm{Stacey} \sur{Alberts}}

\author[3]{\fnm{Fengwu} \sur{Sun}}

\author[1]{\fnm{Jakob M.} \sur{Helton}}

\author[1]{\fnm{Marcia} \sur{Rieke}}

\author[4]{\fnm{Irene} \sur{Shivaei}}

\author[5,6]{\fnm{Francesco} \sur{D'Eugenio}}

\author[5,6]{\fnm{Sandro} \sur{Tacchella}}

\author[7]{\fnm{Brant} \sur{Robertson}}

\author[1]{\fnm{Yongda} \sur{Zhu}}

\author[5,6,8]{\fnm{Roberto} \sur{Maiolino}}

\author[9]{\fnm{Andrew J.} \sur{Bunker}}

\author[1]{\fnm{Yang} \sur{Sun}}

\author[1]{\fnm{Christopher N. A.} \sur{Willmer}}

\affil*[1]{\orgdiv{Department of Astronomy}, \orgname{University of Arizona}, \orgaddress{\street{933 N. Cherry Avenue}, \city{Tucson}, \postcode{85721}, \state{Arizona}, \country{USA}}}

\affil[2]{\orgname{NSF’s National Optical-Infrared Astronomy Research Laboratory}, \orgaddress{\street{950 N. Cherry Avenue}, \city{Tucson}, \postcode{85719}, \state{Arizona}, \country{USA}}}

\affil[3]{\orgname{Center for Astrophysics, Harvard \& Smithsonian},
\orgaddress{\street{60 Garden Street}, \city{Cambridge}, \postcode{02138},
\state{Massachusetts}, \country{USA}}}

\affil[4]{\orgname{Centro de Astrobiolog\'{i}a (CAB), CSIC-INTA}, 
\orgaddress{\street{Carretera de Ajalvir km 4, Torrej\'{o}n de Ardoz}, \city{Madrid}, \postcode{28850}, \country{Spain}}}

\affil[5]{\orgname{Kavli Institute for Cosmology, University of Cambridge},
\orgaddress{\street{Madingley Road}, \city{Cambridge}, \postcode{CB3 0HA}, \country{UK}}}

\affil[6]{\orgname{Cavendish Laboratory, University of Cambridge},
\orgaddress{\street{19 JJ Thomson Avenue}, \city{Cambridge}, \postcode{CB3 0HE}, \country{UK}}}

\affil[7]{\orgname{Department of Astronomy and Astrophysics, University of California, Santa Cruz},
\orgaddress{\street{1156 High Street}, \city{Santa Cruz}, \postcode{95064}, \state{California}, \country{USA}}}

\affil[8]{\orgname{Department of Physics and Astronomy, University College London},
\orgaddress{\street{Gower Street}, \city{London}, \postcode{WC1E 6BT}, \country{UK}}}

\affil[9]{\orgname{Department of Physics, University of Oxford},
\orgaddress{\street{Denys Wilkinson Building, Keble Road}, \city{Oxford}, \postcode{OX1 3RH}, \country{UK}}}


\abstract{
A major unsolved problem in galaxy evolution is the early appearance of massive quiescent galaxies that no longer actively form stars only $\sim$1 billion years after the Big Bang. Their high stellar masses and extremely compact structure\citep{Ji2024} indicate that they formed through rapid bursts of star formation\citep{Merlin2012, Chiosi2014} between redshift $z\sim6-11$ \citep{Glazebrook2023, deGraaff2024, Nanayakkara2024,Weibel2024}. Theoretical models of galaxy evolution cannot  explain their high number density, rapid growth and truncation of star formation at such early times \citep{Gould2023, Valentino2023}, which likely requires extreme feedback to destroy the cold interstellar medium (the fuel for star formation).
We report the discovery of a significant reservoir of hot dust in one of the most distant known examples at $z=4.658$, GS-9209 \citep{Carnall2023}. The dust was identified using JWST's Mid-Infrared Instrument (MIRI), whose unprecedented sensitivity and high spatial resolution, for the first time, firmly show that this dust is significantly more extended than the stars by $\gtrsim3$ times. We find that the dust has preferentially been evacuated or diluted in the galaxy center. Our analysis finds that the extended hot dust emission is consistent with recent heating by a younger and more spatially extended generation of star formation. This reveals that the earliest quiescent galaxies did not form in a single rapid burst; instead, similar to galaxy growth at later times, the center formed first with star formation continuing in an extended envelope. The growth of this galaxy is truncating from the inside out, consistent with central gas depletion from early AGN feedback.
}

\maketitle

The galaxy presented here, GS-9209, is so far one of the earliest known massive quiescent galaxies in the Universe with a spectroscopically confirmed redshift of $z=4.658$  \citep{Carnall2023}. JWST's NIRCam observations \citep{Ji2024} revealed that the rest-optical/NIR half-light radius of GS-9209 is only $r_e\sim200$ pc, which corresponds to a stellar mass surface density of $\sim10^{11}M_\odot/\rm{kpc}^{2}$ approaching the maximum allowed density set by stellar feedback \citep{Grudic2019}. The formation of such an extremely compact stellar morphology, and its connection to the very early truncation of star formation pose major challenges to models of massive galaxy formation in the early Universe.

In December 2022, we obtained 8-band imaging of GS-9209 at observed wavelengths \lamobs\ $\sim5-30\mu m$ with the Mid-Infrared Instrument \citep{miri} (MIRI) of the James Webb Space Telescope \citep{jwst} (JWST) through the SMILES program \citep{Rieke2024,Albertsmiles}. With the MIRI observations, for the first time in a massive quiescent galaxy at this redshift, we are able to constrain emission red-ward of the stellar bump at rest-frame \lamrest\ $=1.6\mu m$ \footnote{Because H$^-$ ion has the minimal opacity at this wavelength, the emission of cool stars shows a maximum/bump in galaxies' SEDs.}. This allows us to put new constraints that were entirely missed by previous observations – the presence of dust emission from GS-9209.

\begin{figure}
\centering
\includegraphics[width=1\textwidth]{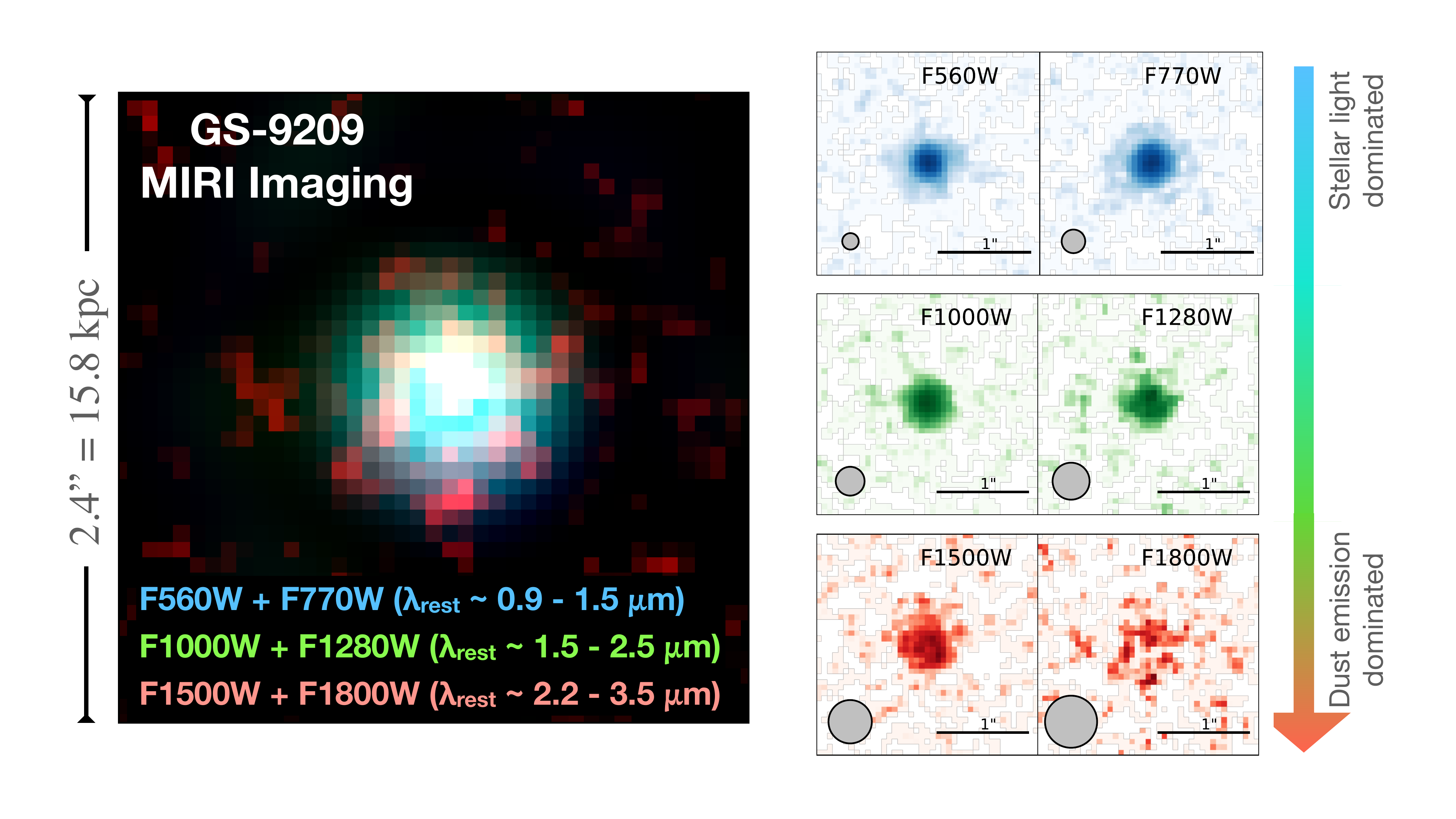}
\caption{{\bf MIRI imaging of GS-9209.} Images shown here are F560W, F770W, F1000W, F1280W, F1500W, F1800W bands that cover \lamobs\ $\sim$ 5.6, 7.7, 10.0, 12.8, 15.0 and 18.0 $\mu m$, respectively. On the left we show the pseudo-RGB image made of the PSF-matched MIRI images. On the right we show the images, in the nominal angular resolution, of individual MIRI bands. From F560W to F1800W, the MIRI imaging samples the spectral range from \lamrest\ $\sim1\mu m$ to \lamrest\ $>3\mu m$, where the contribution from the dust emission becomes increasingly important relative to the starlight.}\label{fig:rgb_miri}
\end{figure}

The MIRI images of GS-9209 are shown in Fig. \ref{fig:rgb_miri}, where the \lamrest\ range probed by each band is indicated. At \lamobs\ $\sim 6-10\mu m$ which corresponds to \lamrest\ $\sim1-2\mu m$, MIRI images show a very compact morphology, suggesting a nucleated, high-density stellar distribution of GS-9209, which is fully consistent with the studies using shorter wavelength NIRCam observations \citep{Carnall2023, Ji2024}.  

\begin{figure}
\centering
\includegraphics[width=1\textwidth]{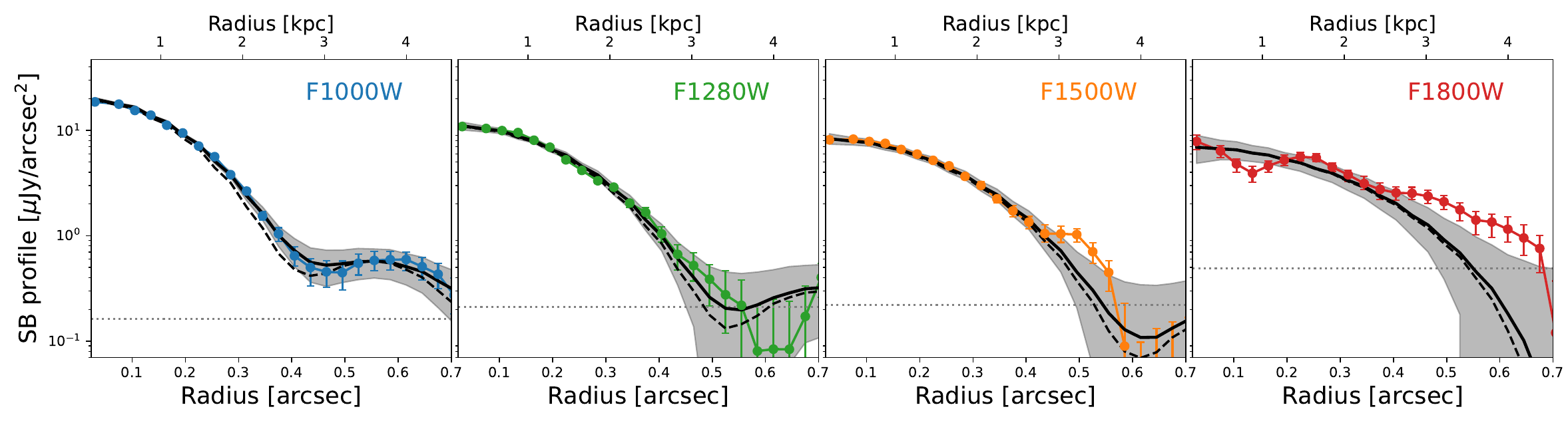}
\caption{{\bf Radial surface brightness profiles of GS-9209 in JWST/MIRI bands at \lamobs\ $>10\mu m$.} In each panel, the observed surface brightness profile is shown as colored points with error bars. The horizontal grey dotted line marks the 3$\sigma$ detection limit. The black solid line and shaded region show the median and 1$\sigma$ of the expected surface brightness profile from our source injection simulations, assuming the galaxy's total light distribution at \lamobs\ $>10\mu m$ were the same as its stellar light. Similarly, the black dashed line shows the expected surface brightness profile for a point source in MIRI. At \lamobs\ $>15\mu m$, corresponding to $\lambda_{\rm{rest}}\gtrsim2.5\mu m$, the light distribution (hot dust emission) of GS-9209 is more extended than its stellar light.}\label{fig:sb_miri}
\end{figure}

Remarkably, moving towards longer wavelengths of \lamobs\ $>10\mu m$, MIRI imaging starts to show more extended emission from GS-9209 than appears at shorter wavelengths. Using source injection simulations, we quantitatively compare GS-9209's light distributions at \lamobs\ $>10\mu m$ to its starlight probed by F770W (see Methods). As Fig. \ref{fig:sb_miri} shows, the F1500W (\lamrest\ $\sim 2.7\mu m$) starts to significantly depart from the stellar profile, a tendency that becomes clearer in both the image and radial profile at F1800W (\lamrest\ $\sim 3.2\mu m$). The confidence level that the light profiles are significantly different is $p>99.99\%$. This difference in light profile is twofold. First, relative to the starlight, F1500W and F1800W images show more extended emission at $0".4<r<0".6$ from the center, with a confidence level of $p=99.06\%$ for F1500W and 99.76\% for F1800W. Second, near the center ($r\sim0".15$) of GS-9209, there appears a flux deficit in the F1800W image, with a confidence level of $p=98.7\%$. As we show in Methods, a similar central flux deficit is also observed in the F2100W image (\lamrest\ $\sim4\mu m$) despite the tentative (S/N $\sim4$) detection of GS-9209 in it.

\begin{figure}
\centering
\includegraphics[width=1\textwidth]{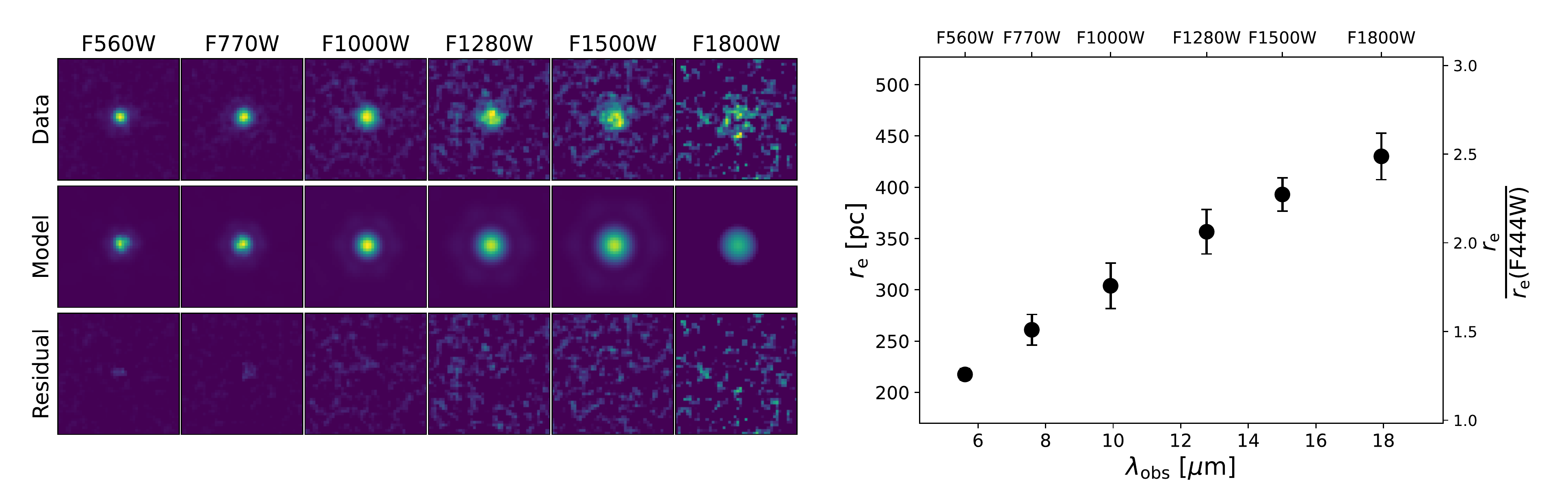}
\caption{{\bf Two-dimensional light profile fitting of GS-9209.} We model the light profile of each one of the MIRI images with a single S\'ersic profile. On the left, we show the comparison between data and best-fit models. On the right, the variation of  $r_{\rm{e}}$ as a function of wavelength is plotted.  The half-light size of the hot dust emission of GS-9209 is significantly more extended than its starlight. 
}\label{fig:galfitm}
\end{figure}

We further study the MIRI light profiles of GS-9209 through two-dimensional parametric light profile fitting. To begin, we assume a single S\'ersic profile \citep{Sersic1963} and simultaneously model the MIRI images from F560W to F1800W (see Methods). As Fig. \ref{fig:galfitm} shows, $r_e$ increases towards longer-wavelength MIRI bands. The $r_{\rm{e}}$ (from the single S\'{e}rsic fitting) of F1800W is $\approx 450$ pc which is $\sim2-3$ times larger than GS-9209's starlight. Motivated by the analysis of the Spectral Energy Distribution (SED)  detailed below, where we find the observed MIRI fluxes at \lamrest\ $ >2.5\mu m$ come from both AGN torus and dust emission of diffuse ISM, we also fit the F1500W and F1800W images using an alternative two-component model, namely a PSF plus a S\'{e}rsic profile. As expected, this two-component fitting returns a larger $r_{\rm{e}}$ of $\approx2$ kpc, almost 10 times of GS-9209's starlight. In Methods, we further test our results by assuming other different light profiles. In all cases the same conclusion is reached that the light distribution of GS-9209 at \lamrest\ $ >2.5\mu m$ is significantly more extended than its starlight, in excellent agreement with our source injection simulations mentioned before. Depending on the assumed light profiles, we estimate the size of GS-9209 at \lamrest\ $ >2.5\mu m$ to be larger than its starlight by at least $\sim2-3$ times, and up to 10 times.

One key revelation from multiple decades of Hubble Space Telescope (HST) \citep{Daddi2005,Damjanov2009,Cassata2011,Cassata2013} and now JWST/NIRCam \citep{Ji2024,Wright2023} observations was that the majority of massive quiescent galaxies at $z>1.5$ have extremely compact stellar morphology.  Moreover, earlier studies found that these galaxies also seem to have rather simple (HST) light profiles \citep{Szomoru2012, Williams2014} that can be well described by a single morphological component.  Yet, our MIRI observations immediately reveal that the structure of GS-9209 is much more complex than we previously thought about typical high-redshift compact quiescent galaxies: Apart from an extremely compact stellar core at the center, GS-9209,  one of the earliest known massive quiescent galaxies, has more extended morphological components, and evidence for substructures (e.g. a near-center flux deficit), at \lamrest\ $\gtrsim2.5\mu m$.

To investigate the physical origin of the MIRI fluxes, we model the SED of GS-9209 (see Methods) using the combined photometry from  the legacy HST/ACS imaging \citep{Giavalisco2004}, the NIRCam imaging from the JADES \citep{Eisenstein2023a} and JEMS \citep{Williams2023} programs, and our new MIRI imaging. Using the integrated photometry (i.e. total observed flux), we obtain quantitatively consistent measures of the stellar-population properties of GS-9209 with those from the SED fitting of \citet{Carnall2023} using NIRSpec spectrum (see Methods).

Crucially, the immense gain of JWST over previous mid-infrared telescopes -- about two orders of magnitude in sensitivity and an order of magnitude in angular resolution compared to Spitzer  \citep{Rieke2024} -- now allows us to perform spatially resolved SED analysis to investigate the detailed internal structure of the earliest massive quiescent galaxies at mid-infrared wavelengths, a knowledge of quiescent galaxies that was totally uncharted beyond the local Universe before JWST.

\begin{figure}
\centering
\includegraphics[width=1\textwidth]{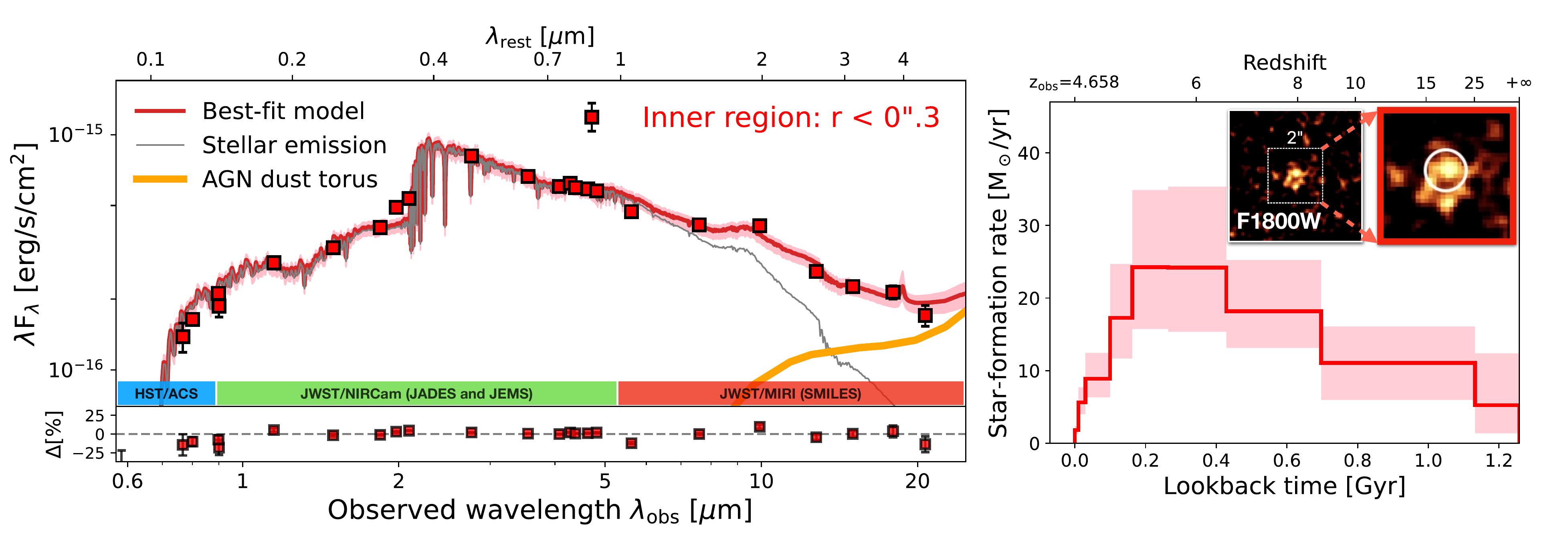}
\caption{{\bf Spectral energy distribution of the inner region ($r<0".3$) of GS-9209.} The left panel shows the comparison between data (squares with error bars) and the best-fit SED model, where the bottom inset shows the relative difference, i.e. $\rm{(Data - Model)/Data}$. MIRI provides key constraints on the emission red-ward the rest-$1.6\mu m$ stellar bump, where the shape of GS-9209's inner SED is inconsistent with the emission from stars alone (grey curve). Hot dust emission from AGN torus (orange) is required to reproduce the MIRI observations. The right panel shows the reconstructed nonparametric star-formation history of the inner region, where solid line and shaded area mark the best-fit and $1\sigma$ uncertainty, respectively. The inset shows the zoomed-in view of GS-9209's F1800W image, with an  $r=0".3$ circular aperture overplotted.
}\label{fig:sed_inner}
\end{figure}

\begin{figure}
\centering
\includegraphics[width=1\textwidth]{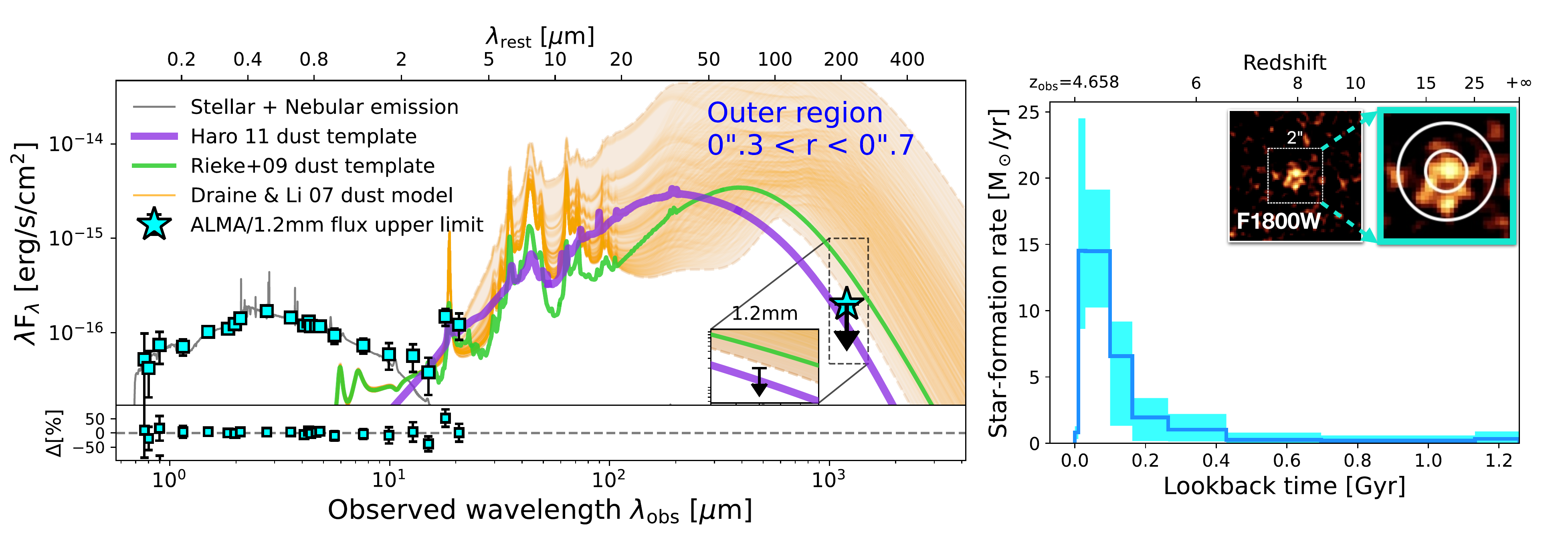}
\caption{{\bf Spectral energy distribution of the outer region ($0".3<r<0".7$) of GS-9209.} Similar to Fig. \ref{fig:sed_inner}, the left and right panels show the best-fit SED model and reconstructed star formation history. Hot dust emission of diffuse ISM is required to reproduce the observed MIRI fluxes. Different dust templates extensively used in the literature are assumed, including \citet{Draine2007, Rieke2009}, and the one of Haro 11 \citep{Lyu2016}. The ALMA non-detection (based on archival data) of GS-9209 suggests that the dust emission of GS-9209 is more consistent with that of Haro 11.
}\label{fig:sed_outer}
\end{figure}

We present the fiducial SED fitting results of GS-9209's inner region enclosed by an $r=0".3$ circular aperture  in Fig. \ref{fig:sed_inner}, and of GS-9209's outer region enclosed by an $r=0".3-0".7$ circular annulus aperture in Fig. \ref{fig:sed_outer}.

GS-9209's outer region is younger and more dust-attenuated than its inner region.  For the inner region, the best-fit model has little instantaneous star-formation activity with a specific star formation rate of $\log \rm{(sSFR/yr^{-1})} = -10.0_{-0.4}^{+0.2}$, a mass-weighted stellar age of $t_{\rm{age}} = 500\pm100$ Myr or a formation redshift of $z_{\rm{form}} = 6.8\pm0.5$, and low dust attenuation with $\rm{E(B-V) = 0.16\pm0.01}$ mag. Comparatively, the outer region has younger stellar populations with a larger $\log \rm{(sSFR/yr^{-1})} = -9.5_{-0.2}^{+0.2}$ and smaller $t_{\rm{age}} = 160\pm60$ Myr or $z_{\rm{form}} = 5.1\pm0.1$, and higher dust attenuation with $\rm{E(B-V) = 0.47\pm0.08}$ mag. In Methods, we test our results by altering the default SED assumptions (star-formation history, metallicity etc.). We show that our conclusions are insensitive to the assumptions made in the fiducial SED fitting.

The spatial difference in stellar populations immediately shows that the formation of GS-9209 did not occur in a single episode. Instead, its formation is in an inside-out manner, which is similar to star-forming galaxies, but with a much higher efficiency since its major star formation has already halted merely $\lesssim 1$ Gyr after the Big Bang.

Starting from \lamobs\ $\sim10\mu m$ or \lamrest\ $\sim2.5\mu m$, the SED of GS-9209 cannot be explained by  purely stellar light alone. The dust emission is required to reproduce the observed MIRI fluxes, for both the inner and outer regions. 

For the inner region, SED analysis suggests that the emission at \lamrest\ $\sim2.5\mu m$ is dominated by AGN torus, a conclusion that is independent of assumed AGN templates (see Methods). We stress that the reprocessed emission of dust associated with diffuse ISM is also included to the SED modeling. Yet, the fitting still finds that the mid-infrared SED of the inner region is  consistent with being dominated by AGN, suggesting the lack of significant amount of dust associated with diffuse ISM at the center of GS-9209 (Fig. \ref{fig:sed_inner}). It is also worth mentioning that, SED fitting using only the shorter wavelength (no MIRI) photometry was ambiguous about the AGN presence in GS-9209 \citep{Ji2024}, although the presence of a faint one is demonstrated by the broad H$\alpha$ line \citep{Carnall2023}. The detection  in our full set of photometry, and the consistency of our SED-inferred AGN luminosity and that  derived from NIRSpec spectroscopy (see Methods) show the power of MIRI in identifying and characterizing AGNs just with photometry. 

For the outer region of GS-9209, we find that the \lamrest\ $>2.5\mu$m emission is observed on kpc-scales, exceedingly larger than the typical (sub-)pc scale of the AGN-heated torus emission at these wavelengths \citep{Koshida2014, Lyu2019}. This suggests that
the dust emission has to be associated with diffuse ISM. Regardless of assumed dust templates, however, if we adopt the energy balance criterion where all starlight attenuated by the dust is re-emitted in infrared, we find that the MIRI fluxes at \lamrest\ $\sim2.5\mu m$ is not well-modeled (see Methods). In fact, earlier studies have questioned the use of the energy balance assumption especially when the UV and IR emission of galaxies is offset from one another  \citep{Casey2017}. Given the significant difference in GS-9209's stellar and dust morphologies revealed by our MIRI imaging, the ineffectiveness of the energy balance is arguably expected for GS-9209. 

Without the assumption of energy balance, as Fig \ref{fig:sed_outer} shows, different dust emission templates are all able to fit the HST-to-MIRI photometry reasonably well. The non-detection of GS-9209 from archival ALMA observations \citep{Hatsukade2018} provides additional, critical constraints on its dust properties. We derive an upper limit at observed 1.2 mm (or \lamrest $\sim210\mu$m) of 80 $\mu$Jy with forced photometry following \citet{Williams2024}. This limit is inconsistent with typical dust models assumed for massive star-forming galaxies \citep[][]{Draine2007, Rieke2009}. Instead, it is broadly consistent with the dust emission template of Haro 11, a starbursting dwarf galaxy in the local Universe whose IR SED -- relative to typical star-forming galaxies -- features a higher dust temperature ($\sim$47 K) and an excess of near- to mid-IR continuum emission likely from additional even warmer dust \citep{Lyu2019}. GS-9209 and Haro 11 share two major similarities, namely a low metallicity and very compact morphology, which indeed have been shown to lead to a warmer/bluer IR SED similar to that of Haro 11 \citep{DeRossi2018}. Future spectroscopy at mid-to-far infrared wavelengths is required to further quantify the dust properties in such systems.

Combining together our morphological and SED analysis, we conclude that there is a significant dust reservoir in GS-9209. The hot dust emission has two origins. At the center, the emission is dominated by AGN torus. In the outskirts of GS-9209 where its stellar population is younger than the inner region, the extended \lamrest\ $>2.5\mu m$ emission is associated with the hot dust of diffuse ISM, mostly likely powered by rest-frame UV/optical starlight absorbed by the dust and re-emitted in IR.

The discovery that the first massive quiescent galaxies like GS-9209 form and quench from the inside out now enables new empirical constraints on their formation pathways. The structure of GS-9209, namely a compact older stellar core surrounded by a younger stellar and dust envelope, is very similar to the hallmark prediction for the gas-rich processes associated with galaxy formation  \citep{Dekel2014}. Physically, the formation of the extremely dense stellar core requires highly dissipative gas accretion \citep{Danovich2015}. The buildup of the core in turn will deepen the gravitational potential well, making the later, continuous accretion of gas and satellite mergers settle into orbits surrounding the older central core. Interestingly, hydrodynamical simulations with adequate resolution for the modeling of relevant gaseous astrophysics predict that continuous gas accretion (while star-formation at the center is effectively halted) should form a extended, ring structure in massive galaxies at a physical scale of $r\gtrsim3\,r_{\rm{e}}^{\rm{star}}$ at $z<4$ \citep{Tacchella2016, Dekel2020,Lapiner2023}, in quantitative agreement with the scale where the extended hot dust emission of GS-9209 is observed. Thus, the gas-rich processes found at $z<4$ may plausibly extend to higher redshifts for the formation of this earlier generation of massive galaxies \citep{Ji2024}.

Near the center, the likely flux deficit at \lamrest\ $>2.5\mu$m (Fig. \ref{fig:rgb_miri}, \ref{fig:sb_miri}), and the lack of evidence of the presence of dust associated with diffuse ISM  (Fig. \ref{fig:sed_inner}) imply central gas depletion as a direct cause of the inside-out quenching of GS-9209. A natural consequence of highly dissipative gas accretion is also the faster, more efficient growth of supermassive black holes \citep{Lapiner2021,Byrne2023}. The impact of this is that AGN feedback may become energetic enough to impact the host galaxy on faster timescales, helping to heat up and even expel gas. GS-9209 indeed hosts an AGN, which indicates that the AGN feedback might be the cause of the vacated dust/gas near the center. 

We finally discuss our MIRI discovery of the extended hot dust emission from GS-9209 in the cosmological context. Numerous studies have now noted that state-of-the-art cosmological simulations fail to quantitatively reproduce the observed number density of massive quiescent galaxies that form at $z>3$ \citep{Pathak2021,Hartley2023, Gould2023, Valentino2023}. A common attribute among these simulations is that the AGN feedback, in particular the black hole seeding time and the time at which kinetic quasar-mode feedback becomes influential, does not occur early enough in cosmic time\citep{Park2023, KurinchiVendhan2023, Kimmig2023}. This has been identified as a likely cause why quenching in simulations occurs too late. The discovery by JWST of abundant supermassive black holes earlier and more massive than expected \citep{Maiolino2023,Stone2024} further corroborates this evidence that AGN feedback is more prevalent and impacts their host galaxies earlier than previously thought. If GS-9209 is a good representative of the general population of earliest massive quiescent galaxies, our MIRI observations provide strong evidence that the gas-rich dissipative processes and AGN feedback play a critical role in the formation and quenching of the earliest quiescent galaxies discovered by JWST. Simulations of sufficiently massive halos at $z > 5$ with the resolution to adequately model gas-rich galaxy growth do not currently exist, but this advance is needed to determine whether the detailed physical properties of galaxies like GS-9209, such as the multi-wavelength structure, are in tension with cosmological expectations.

With its revolutionary capabilities, particularly the unprecedented angular resolution, JWST's MIRI now opens a new window of studying high-redshift, freshly quenched galaxies through the spatial distribution of dust emission, promising powerful constraints on the physical mechanisms responsible for the cessation of star formation in the earliest massive quiescent galaxies. 

\section*{Methods}\label{method}

\section{Cosmology model and definitions}

Throughout this study, we adopt the $\Lambda$CDM cosmology with $H_0$ = 70 km s$^{-1}$ Mpc$^{-1}$, $\Omega_{\rm m}$ = 0.3 and $\Omega_{\rm \Lambda}$ = 0.7. All magnitudes are presented in the AB system.

\section{NIRCam and MIRI imaging data reduction}

GS-9209 appears in the medium-depth portion of the JADES program in GOODS-South \cite{Eisenstein2023a, Eisenstein2023b}, where we have obtained JWST/NIRCam imaging in 8 bands, F090W, F115W, F150W, F200W, F277W, F356W, F410M and F444W. Additional 5 medium-band NIRCam images, i.e. F182M, F210M, F430M, F460M and F480M, were obtained by the JEMS program \citep{Williams2023}. We also include public, shallower F182M, F210M and F444W imaging from the FRESCO survey \citep{Oesch2023} to further improve the depth of those images. 
We reduce all NIRCam imaging observations using the pipeline developed by JADES, the details of which have been presented in \citet{Rieke2023}. In brief, we process the raw images using the JWST Calibration Pipeline with the following custom corrections. During Stage 1 of the JWST pipeline, we mask and correct for the “snowballs”\footnote[2]{\href{snowball}{https://jwst-docs.stsci.edu/jwst-near-infrared-camera/nircam-instrument-features-and-caveats/nircam-claws-and-wisps}} cosmic-ray artefacts. During Stage 2 of the JWST pipeline, we remove the $1/f$ noise and subtract 2D background from the images, and also correct the “wisp” scattered light features. Afterwards, we tie the astrometry of individual exposures of a given visit to the World Coordinate System (WCS) of a reference catalog constructed from HST/WFC3 F160W mosaics in GOODS-South with astrometry tied to Gaia-EDR3 \citep{Gaia2021}. Before combining visit-level images to create the final mosaic, we manually inspect all individual exposures containing GS-9209. We remove two out of six F090W exposures from our analysis as the ``snowballs" effect is seen in very close proximity ($<2"$) of GS-9209 and significantly affects the surrounding sky.

GS-9209 also appears in the JWST/MIRI program, SMILES \citep{Rieke2024,Albertsmiles}, where we have obtained MIRI images in 8 bands, F560W, F770W, F1000W, F1280W, F1500W, F1800W, F2100W and F2550W. We reduce MIRI imaging observations using the nominal JWST Pipeline with the latest JWST Calibration Reference System. We make custom modifications for sky subtraction which has been described in detail in \citet{Albertsmiles}. In brief, to better correct for the large spatial gradient of sky background in our data and remove substructures including tree-ring shaped features along the columns and rows of the MIRI detector, for each one of the cal files (i.e., the products of the {\sc calwebb\_image2} in JWST Pipeline), we adopt an iterative process to (1) progressively mask sources, (2) median filter out large-scale sky gradients and striping features along detector columns and rows and (3) construct a super background by median combining and scaling all those filtered cal files. Finally, we subtract from each cal file the super background, after which we do an additional median subtraction using a 256$\times$256 pixel$^2$ box to remove remaining background variations, which are mainly caused by cosmic ray showers. We tie the astrometry of  MIRI mosaics to be the same as JADES.

\begin{figure}
    \centering
    \includegraphics[width=1\textwidth]{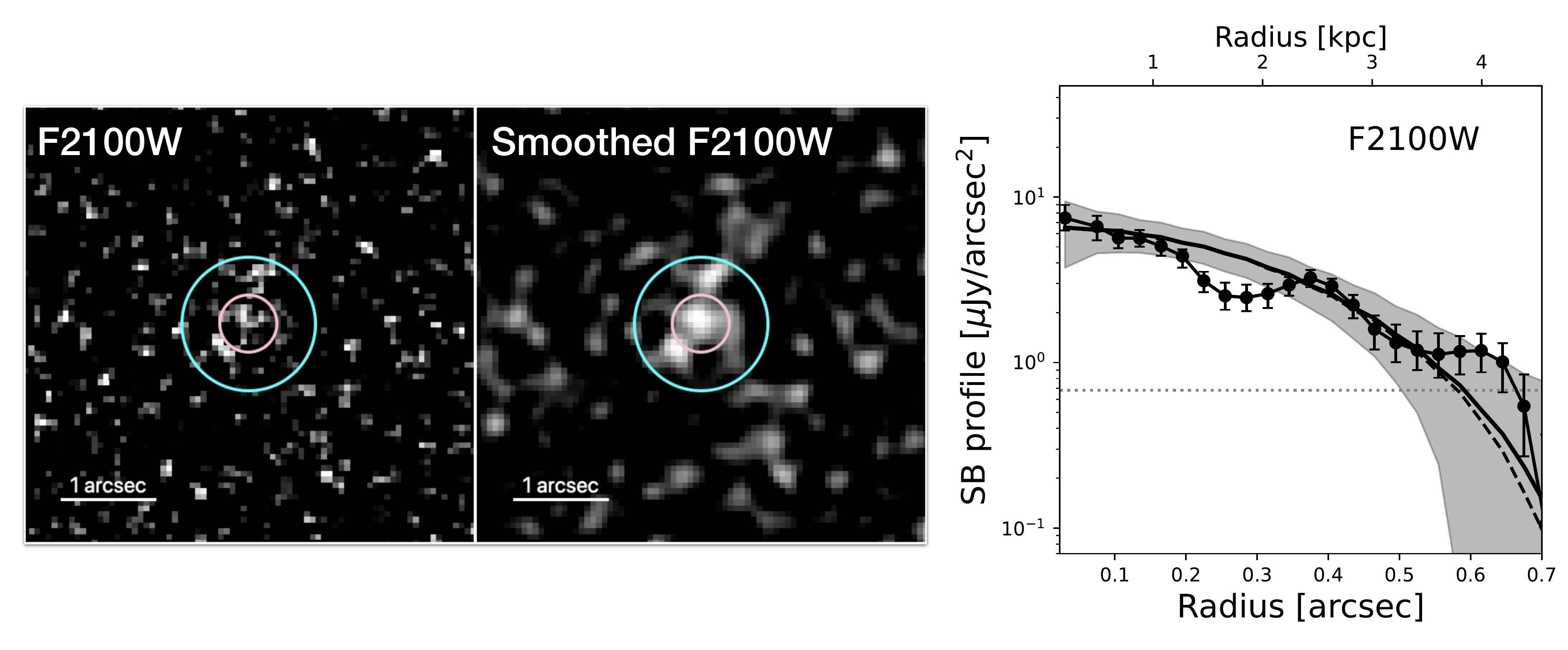}
    \caption{MIRI/F2100W observations of GS-9209. The left panel shows the original F2100W image. The middle panel shows the F2100W image smoothed with a $\sigma=1$ pixel Gaussian kernel. The pink and cyan circles have a radius of 0".3 and 0".7, respectively. The right panel shows the result of source injection simulations, similar to those shown in Fig. \ref{fig:sb_miri}.}
    \label{fig:21um}
\end{figure}

Current imaging depth of MIRI/F2100W from SMILES results in a signal-to-noise ratio S/N $\sim4$ detection of GS-9209 (Fig. \ref{fig:21um}). For the reddest MIRI band F2550W, GS-9209 is not detected (S/N $<1$) which however is expected based on the detection limit of SMILES. The results of these two band MIRI images are not included to the main text.

\section{Analysis of MIRI light profiles} \label{method:miri_sb}

\subsection{Source injection simulations}

The main goal of this analysis is to compare observed light distributions of GS-9209 at \lamobs\ $>10\mu m$ with its stellar light distribution. A common method is through parametric morphology modeling, e.g. S\'{e}rsic fitting, to compare the fitted parameters at different wavelengths. We stress, however, that the spatial distribution of hot dust emission in high-redshift quiescent galaxies is largely unknown at present.  Thus, the mismatch between the assumed parametric light profile and the intrinsic distribution of hot dust emission can make such comparison fraught with systematic errors. To mitigate this issue, we decide to make the comparison through a purely empirical method -- source injection simulations. 

The idea is to inject artificial sources with the same light distribution of GS-9209's starlight (F770W) into empty regions of MIRI images at $>10\mu m$, and then to measure surface brightness profiles of these injected sources with the same procedure used for GS-9209. If the light distributions of GS-9209 at \lamobs\ $>10\mu m$ differ from its starlight distribution, then we expect the observed surface brightness profiles of GS-9209 at \lamobs\ $>10\mu m$ to be different from the recovered profiles of the realizations (i.e. injected artificial sources). 

For source injections, we first create artificial galaxy images mimicking the stellar light distribution of GS-9209. The typical light distribution of $z>3$ quiescent galaxies at rest-frame NIR remains largely unknown, but can be complex \citep{Ji2024}. Thus, instead of assuming a parametric, likely over-simplified profile of GS-9209's starlight distribution, we directly use its MIRI/F770W image to generate synthetic images at longer wavelengths. The choice of F770W is because (1) GS-9209 is detected in F770W at high S/N ($=72$, using a $r=0.25"$ circular aperture), (2) AGN contribution is negligible at F770W based on multiple different analyses of GS-9209's SED, and (3) at GS-9209's redshift $z=4.658$, F770W probes \lamrest $=1.0\sim1.7\mu m$ -- covering the 1.6$\mu m$ stellar bump -- which is a very sensitive to (and arguably the best available probe of) the spatial distribution of stars in GS-9209. 

We PSF-match the F770W image to the MIRI bands at \lamobs\ $>10\mu m$. We rescale the PSF-matched F770W image of GS-9209 such that the flux within a $r=0.2"$ circular aperture is matched to that observed at longer wavelengths. We then use the MIRI segmentation map to mask out all detected sources, and inject the rescaled synthetic images to the $1'\times2'$ ($\sim$ the field of view of MIRI imaging) region centered around GS-9209. We randomly inject these images $10^4$ times, and each time we also resample the synthetic image with the Poisson noise calculated using the corresponding exposure time maps of our MIRI observations. Finally, we measure the surface brightness profiles of the injected artificial sources. The median and $1\sigma$ range of the surface brightness profiles of the $10^4$ realizations are shown as black solid line and shaded region in Fig. \ref{fig:sb_miri}. In addition to injecting artificial sources with the observed F770W light profile, we also run another set of simulations to inject entirely unresolved point sources. In Fig. \ref{fig:sb_miri}, the median surface brightness profiles of the injected artificial point sources are shown as black dashed line. 

Before moving forward, we stress that, in contrast to simply comparing the PSF-matched F770W image of GS-9209 to its images at longer wavelengths, source injection simulations take into account the realistic noise behaviors of the MIRI observations, allowing us to have a robust estimate on the significance of the difference in light profiles between the stellar light and hot dust emission of GS-9209.

We first test the null hypothesis that the observed light distributions of GS-9209 at \lamobs\ $>10\mu m$ are similarly compact to its stellar light distribution. Out of $10^4$ realizations, only in 94 and 24 cases is the surface brightness profile of injected artificial sources as bright or brighter than that observed in F1500W and F1800W within a $0".4<r<0".7$ annulus aperture. We thus reject the null hypothesis with a p-value of $99.06\%$ and 99.76\% for the F1500W and F1800W images, respectively. This shows that the observed MIRI light distributions of GS-9209 at \lamrest\ $>2.5\mu m$ (hot dust emission) are more extended than its stellar light. We note that, despite GS-9209 only being detected with a moderate significance in MIRI/F2100W, source injection simulations were still preformed for F2100W and the results are shown in the right panel of Fig. \ref{fig:21um}. Owing to the relatively low S/N, the observed F2100W light profile at $0".4<r<0".7$ is within $1\sigma$ range of the source injection simulations, although the observed one is above the median surface brightness profile measured from the simulations, i.e. in line with the results of F1500W and F1800W.

Relative to the stellar light distribution, there appears to be a flux deficit near the center ($r\sim0".15$) of GS-9209 in its F1800W image (Fig. \ref{fig:rgb_miri} and \ref{fig:sb_miri}). Such an angular scale is smaller than the resolution of MIRI/F1800W imaging (FWHM/2 $\sim$ 0".25), making it very difficult, if at all possible, to quantify precisely the physical scale over which the flux deficit is present. Nonetheless, with source injection simulations, we can at least test the null hypothesis that this deficit is a result of noise fluctuations. Out of $10^4$ realizations, the surface brightness profiles of 130 injected artificial sources are as faint or fainter than that observed in F1800W within a $0".12<r<0".18$ annulus aperture. We thus reject the null hypothesis with a p-value of 98.7\%. We note that, at a similar angular scale ($0".15<r<0".3$), a flux deficit is also observed in the F2100W image (Fig. \ref{fig:21um}). We thus believe this deficit to be a real feature of GS-9209's hot dust emission. 

Finally, among F1800W surface brightness profiles of the $10^4$ injected artificial sources, none of them has both the near-center flux deficit at $r\sim0".15$ and more extended emission at $0".4<r<0".7$. Thus, the stellar and hot dust distributions of GS-9209 are different at a significance level of $>99.99\%$. We stress again that we intend to make the source injection simulations as empirical as possible such that our conclusions have the least dependence, if at all, on the assumptions made. We conclude that the hot dust distribution of GS-9209 differs significantly from its starlight.

\subsection{Parametric light-profile fitting}

For high-redshift quiescent galaxies, we reiterate that the spatial distribution of hot dust emission is unknown. Nonetheless, a meaningful estimate of the difference between the stellar and hot dust distributions is still very important, as this will enable quantitative comparisons between the MIRI observations and predictions from hydrodynamical simulations. In what follows, we thus use a few commonly assumed parametric light profiles to model the MIRI images, in an attempt to quantify the difference in size between the stellar and hot dust emission of GS-9209.

The analysis is done by performing PSF-convolved fitting of the 2D light distributions from F560W to F1800W. We use the software {\sc Galfitm} \citep{Haussler2013,Haussler2022} that is built upon {\sc Galfit3} \citep{Peng2010} but also allows simultaneously modeling multi-band images. For PSF models of MIRI images at \lamobs\ $>10\mu m$, i.e. from F1000W to F1800W, there is a very limited number of MIR-bright point sources in the SMILES footprint, leading to noisy outskirts of the empirical PSF models. To mitigate this, instead of using the empirical PSFs, we build model PSFs at \lamobs\ $>10\mu m$ following the strategy  used for the NIRCam imaging of JADES \citep{Ji2023}. In short, we first inject the {\sc Webbpsf} \citep{Perrin2014} models into the individual stage-2 images, and then mosaic them following the same stage-3 data reduction of SMILES. The PSF models are then constructed from these PSF-mosaics. The empirical and model PSFs agree with each other very well at the center, but model PSFs have a much more stable behavior at the outskirts. For F560W and F770W, we do not use model PSFs, because the cross-shaped artifact (a.k.a. cruciform), an extended artifact caused by internal reflection in the F560W and F770W detectors \citep{Gaspar2021}, is not well captured in current {\sc Webbpsf} modeling. Instead, for F560W and F770W we use the empirical PSF models built upon observed point sources in SMILES following the method of \citet{Anderson2010}. 

To begin, we first assume a single S\'ersic profile for each one of the MIRI images. By default, because the S/N of the detections decrease with wavelength, we simultaneously model all MIRI images such that the fit uses the information of all the available data while allowing parameters to vary as a smooth function of wavelength. Such an approach has been demonstrated to provide more stable and accurate multi-wavelength morphological parameters than modeling images of different bands independently, particularly when some images have  low-to-moderate S/N  \citep{Haussler2013,Dimauro2018,Nedkova2021,Haussler2022}. During the fitting, we fix the sky background to the 3$\sigma$-clipped median pixel value of a $5"\times5"$ cutout after masking all detected sources using the segmentation map. 
We fit the center, axis ratio and position angle as free parameters, but assume they do not vary with wavelength. We fit the S\'ersic index $n$ and allow it to change freely with wavelength. The main parameter of interest, the effective radius $r_{e}$, is assumed to vary with wavelength following a second order Chebyshev polynomial, following recent studies \citep{Nedkova2021}. To derive the uncertainties of the fitted parameters, we use the error flux extension of the MIRI images to Monte Carlo resample the image pixel values, and then run {\sc Galfitm}. We repeat the procedure 1000 times, and use the standard deviation as the parameter uncertainties. The fitting results are shown in Fig. \ref{fig:galfitm}. We obtain $r_{e}= 217\pm10$ pc in F560W and $r_{e}= 261\pm17$ pc in F770W, in good agreement with the size measured from the NIRCam observations within the uncertainty \citep{Carnall2023, Ji2024}. For F1800W, we obtain $r_{e}= 430\pm27$ pc. 

We test the single S\'{e}rsic fitting results above by modeling each one of the MIRI images independently, namely that, except fixing the centroid, all other parameters, axis ratio, position angle, S\'ersic index $n$ and $r_{e}$ are allowed to vary freely with wavelength. We obtain $r_{e}= 216\pm12$, $225\pm17$ and $461\pm49$ pc in F560W, F770W and F1800W, respectively. We also test our results by (1) fitting the sky background as a free parameter and (2) using empirical PSFs (though low S/N at outskirts) at \lamobs\ $>10\mu m$. We do not see any substantial changes in our results. Therefore, our parametric light-profile fitting reaches to the same conclusion as our source injection simulations: The light distribution of GS-9209 at \lamrest $>3\mu m$ is more extended than its stellar light. Assuming a single S\'{e}rsic profile, the size of hot dust emission of GS-9209 is $\gtrsim 3$ times larger than its stellar light.

The quantitative difference in size between the stellar and hot dust emission depends on the parametric light profile assumed for the modeling. The presence of AGN in GS-9209, which is revealed by both broad H$\alpha$ emission \citep{Carnall2023} and our SED fitting with MIRI, motivates the fit to the MIRI images using an alternative two-component model, a PSF component to account for the AGN, plus a S\'{e}rsic component to account for the host galaxy. During this modeling, each one of the MIRI images is fitted independently. From F560W to F1280W, the PSF+S\'{e}rsic fitting returns essentially the same results as the single S\'{e}rsic fitting: flux contribution from the PSF component is $<10\%$. The lack of significant contribution from the PSF component suggests that a very compact stellar structure dominates GS-9209's emission only up to \lamrest $\sim2 \mu m$, a conclusion independently reached by our SED analysis below. For F1500W and F1800W, flux contributions from the PSF and from the S\'{e}rsic component become comparable. The flux ratio of the PSF to the S\'{e}rsic component is 0.46 and 0.65 for F1500W and F1800W, respectively. This suggests the observed MIRI fluxes at \lamrest $\sim3-5\mu m$ are likely associated with two origins, i.e. AGN dust torus and the dust emission from diffuse ISM, a conclusion again independently reached by our SED analysis. We note, however, that with current F1500W and F1800W imaging, the reduced $\chi^2$ between the single S\'{e}rsic fitting and the PSF+S\'{e}rsic fitting are statistically indistinguishable. 

The PSF+S\'{e}rsic fitting returns $r_e=2.4
\pm0.6$ kpc in F1800W, with the best-fit S\'{e}rsic index of $n\sim0.2$ which hits the minimal $n$ commonly allowed in S\'{e}rsic modeling \citep[][]{vanderwel2012}. We therefore perform another PSF+S\'{e}rsic fitting by fixing $n=1$ (exponential disk), which returns $r_e = 1.8\pm0.5$ kpc in F1800W, i.e. $\approx 10$ times of the stellar $r_e$. We note that, relative to the single S\'{e}rsic fitting, this significant increase in $r_e$ derived from the PSF+S\'{e}rsic fitting is expected, because adding a PSF is equivalent to fitting the F1800W image of GS-9209 with some fraction of central (non-stellar) flux removed. We also note that, unlike the $r_e$ from the single S\'{e}rsic fitting which is a characteristic size of total hot dust emission regardless of its origin, the $r_e$ from the PSF+S\'{e}rsic fitting should be considered as the size of hot dust emission that is only associated with the diffuse ISM of GS-9209.

\section{Analysis of spectral energy distribution}

We perform detailed analysis of the spectral energy distribution (SED) of GS-9209 via SED modeling, with an emphasis on constraining the origin of the observed MIRI fluxes. Data included in the modeling is the photometry of 25 filters from HST/ACS (F435W, F606W, F775W, F814W and F850LP), JWST/NIRCam (F090W, F115W, F150W, F182M, F200W, F210M, F277W, F356W, F410M, F430M, F444W, F460M and F480M) and JWST/MIRI (F560W, F770W, F1000W, F1280W, F1500W, F1800W and F2100W). All photometry is measured using the PSF-matched images where all images are PSF-homogenized to MIRI/F2100W, the longest wavelength band where GS-9209 is detected. During our SED fitting, an error floor is imposed on photometry: the uncertainty of flux is not allowed to be smaller than 5$\%$, the typical uncertainty caused by imperfect flux calibration and PSF homogenization.  

The SED fitting is done mainly with the software {\sc Prospector} \cite{Johnson2021}. The basic setups of our {\sc Prospector} fitting are detailed as follows. We fix the redshift to be $z=4.658$ \citep{Carnall2023}. We use the \citet{Kroupa2001} stellar initial mass function. We adopt the FSPS stellar synthesis code \citep{Conroy2009,Conroy2010} with the stellar isochrone libraries MIST \citep{Choi2016,Dotter2016} and the stellar spectral libraries MILES \citep{FalconBarroso2011}. We use the \citet{Madau1995} IGM transmission model. We include the model of \citet{Byler2017} for nebular emission. By default, we treat differently the dust attenuation towards nebular emission and young ($<10$ Myr) stellar populations, and towards old ($>10$ Myr) stellar populations \citep{Charlot2000}. The dust attenuation of the former is modelled as a power law, while the latter is modelled following the parameterization of \citet{Noll2009}, i.e., a modified \citet{Calzetti2000} dust attenuation law with a Lorentzian-like profile to describe the 2175\AA\ dust feature. Following \citet{Kriek2013}, we tie the strength of the 2175\AA\ feature to the dust index of \citet{Noll2009}.

For the fiducial star formation history (SFH), we use a nonparametric, piece-wise form composed of 9 lookback time bins with the continuity prior \citep{Leja2019}. The first three lookback time  bins are fixed to be $0-10$, $10-30$ and $30-100$ Myr to capture recent episodes of star formation with relatively high time resolution. The last lookback time bin is $0.9t_{\rm{H}} - t_{\rm{H}}$ where $t_{\rm{H}}$ is the Hubble Time at $z=4.658$. The remaining five bins are evenly spaced in logarithmic space between 100 Myr and 0.9$t_{\rm{H}}$. Such a choice of lookback time bins is very similar to those extensively used in recent studies of high-redshift massive galaxies \citep{Tacchella2022,Leja2022,Ji2022a,Ji2023a, Williams2023b}. Nonetheless, in a later Section, we still test our SED fitting results using different time bins of nonparametric SFH.  

With photometric data alone, it remains difficult to tightly constrain stellar and gas-phase metallicities. By default, we thus use strong priors based on the metallicity measures and their uncertainties from NIRSpec spectroscopy of \citet{Carnall2023}. Specifically, for stellar metallicity, we assume a Gaussian prior centered at $0.11Z_\odot$ with a width of 0.1 dex. For gas-phase metallicity, we assume a Gaussian prior centered at $0.17Z_\odot$ with a width of 1 dex. We shall test our SED fitting results using different metallicity assumptions in a later Section.

The MIRI imaging covers the spectral range of GS-9209 towards \lamrest $\sim 3-5\mu m$ where hot dust emission, if present, becomes increasingly important \citep{Draine2003,Conroy2013}. We thus also add models of dust emission to the SED modeling (see below for details). 

\subsection{SED modeling of the entire galaxy} \label{method:sed_tot}

We start by fitting the SED of GS-9209 using integrated photometry, i.e. total flux within a circular aperture of $r=0.7"$ which is about the size of the galaxy's extent in the segmentation map. For dust emission, here we use models of \citet{Nenkova2008a,Nenkova2008b} to model the emission from AGN dust torus, and of \citet{Draine2007} to model the reprocessed emission of dust associated with diffuse ISM. These two models are the default of {\sc Prospector} and have been widely used in the literature.

We do not include the results from this fitting of integrated photometry to the main text, as this analysis is mainly served as a consistency check with the previous inference of GS-9209's stellar-population properties done by \citet{Carnall2023} who performed the SED fitting mainly with the JWST/NIRSpec fixed-slit spectroscopy with medium resolution gratings of G235M and G395M.

\begin{figure}
    \centering
    \includegraphics[width=1\linewidth]{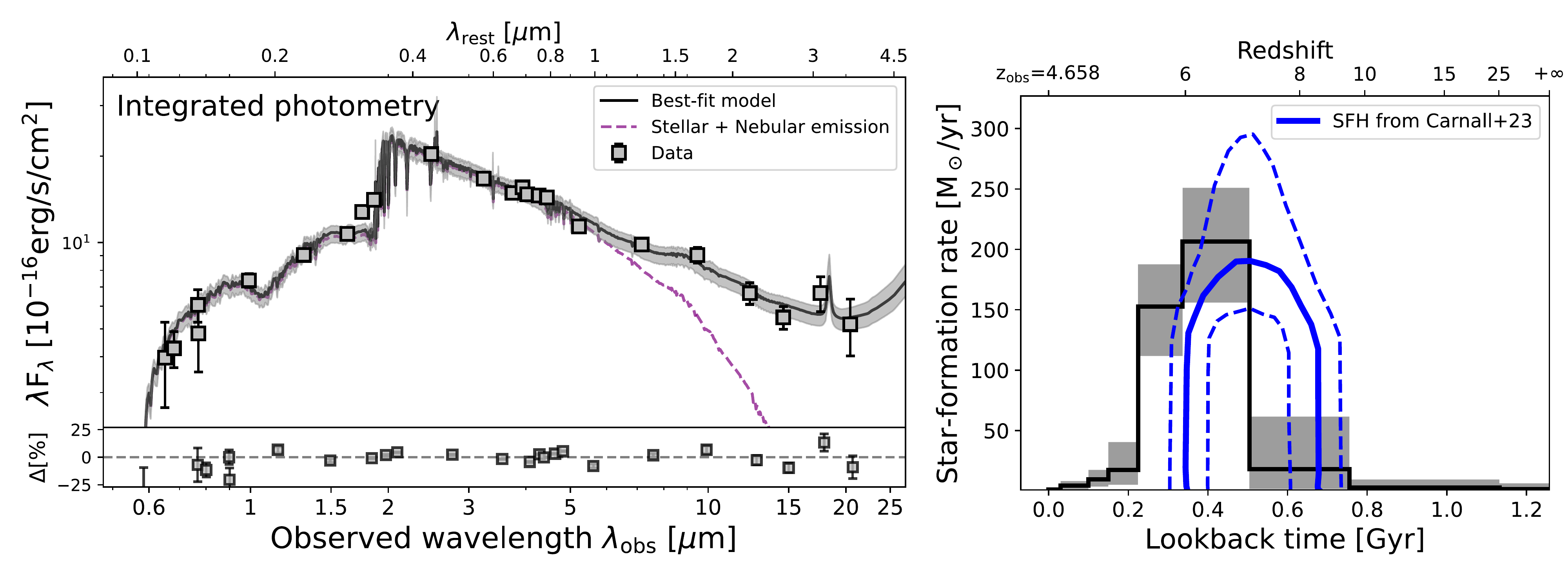}
    \caption{SED fitting of the integrated photometry of GS-9209. The left panel shows the comparison between the best-fit model and data. The magenta dashed line shows the emission from the stellar and nebular emission of GS-9209. The dust emission is clearly required to reproduce the observed MIRI fluxes. The right panel shows the reconstructed SFH of GS-9209. For comparison, we plot in blue the SFH (both the best-fit and 1$\sigma$ uncertainty)  measured by \citet{Carnall2023} using NIRSpec spectroscopy. }
    \label{fig:tot_sed}
\end{figure}

The left panel of Fig. \ref{fig:tot_sed} shows the SED fitting result of integrated photometry. The best-fit model shows a strong rest-frame 4000\AA\ break and prominent absorption features associated with A-type stars, in line with the NIRSpec spectroscopy of \citet{Carnall2023} confirming the quiescent/post-starburst nature of GS-9209. More importantly, with MIRI photometry, it immediately becomes clear that the hot dust emission is required to reproduce the SED of GS-9209 at \lamrest $>2\mu m$. The origin of the hot dust emission will be elaborated in greater detail in following Sections through analysis of the inner and outer regions of GS-9209. 

Using integrated photometry, we obtain a stellar mass of $\log_{10}(M_*/M_\odot)=10.62^{+0.03}_{-0.02}$, in excellent agreement with that ($\log_{10}(M_*/M_\odot)=10.58\pm{0.02}$) measured by \citet{Carnall2023} using NIRSpec spectroscopy. 

We obtain a star formation rate of SFR $ = 2.7^{+3.0}_{-2.4}\, M_\odot / \rm{yr}$. The SFR from SED fitting can be sensitive to the assumed SFH \citep{Conroy2013}. If we use a parametric delayed-tau SFH, i.e. $\rm{SFR}(t)\propto t e^{-t/\tau}$, we would obtain SFR $ = 1.0^{+2.6}_{-0.8}\, M_\odot / \rm{yr}$. We note that our SED-inferred SFR, no matter which SFH is assumed, is larger than that inferred from the SED fitting of \citet{Carnall2023} which suggested that the SFR of GS-9209 is consistent with zero over the past 100 Myr. Interestingly, however, our SFR is actually in excellent agreement with what is inferred based on the narrow component of H$\alpha$ emission (SFR$_{\rm{H}\alpha,\rm{narrow}} = 1.9\pm0.1M_\odot/\rm{yr}$) in the NIRSpec spectrum of \citet{Carnall2023}. We mention that, even with the higher SFR from our measurement, the specific star formation rate (sSFR) of GS-9209 remains to be very low, i.e. $\log_{10}(\rm{sSFR/yr^{-1}})=-10.1^{+0.5}_{-0.4}$ corresponding to $\sim 0.1/t_{\rm{H}}$ which is smaller than the sSFR threshold of $0.2/t_{\rm{H}}$ normally used for identifying quiescent galaxies at high redshift \citep{Carnall2023b}.

Finally, in the right panel of Fig. \ref{fig:tot_sed}, we compare the reconstructed SFH of our measurement with that of \citet{Carnall2023}. Overall, the two SFHs are similar: The major stellar mass assembly of GS-9209 started $\sim500$ Myr ago, i.e. $z\sim7$, with a peak star formation rate SFR$_{\rm{peak}}\sim 200M_\odot/\rm{yr}$. Noticeably, however, the stellar population of GS-9209 inferred from our SED fitting is younger than that from \citet{Carnall2023} by $\sim 100$ Myr (mass-weighted stellar age). We note that NIRSpec's fixed slit (S200A1) only has a width of 0".2, while in this work we find that GS-9209's outskirts are younger than its center. This can explain the younger stellar population found by our SED fitting of integrated photometry ($r=0".7$ aperture). 

\subsection{SED modeling of the inner region} \label{method:sed_inner} 
Here we describe our SED analysis of the inner region of GS-9209. The inner region is enclosed by an $r=0".3$ (similar to the FWHM of MIRI/F2100W image) circular aperture centered at the centroid of F770W, i.e. the stellar continuum of GS-9209. 

For dust emission, by default we use the same models as used for the  SED fitting of integrated photometry (Section \ref{method:sed_tot}). The results have been presented in Fig. \ref{fig:sed_inner} and discussed in detail in the main text. In what follows, we alter our SED modeling significantly to check that if our conclusion about the origin of MIRI fluxes is sensitive to the default SED assumptions.

First, we replace both the default \citet{Draine2007} dust model and the \citet{Nenkova2008a, Nenkova2008b} AGN torus model with the semi-empirical dust emission models of \citet{Lyu2023}. The \citet{Lyu2023} templates are calibrated against various observations both for UV-to-IR AGN emission and for galaxy dust emission associated diffuse ISM \citep{Lyu2022,Lyu2023}. Second, because of the highly uncertain dust extinction of AGN, instead of assuming the same dust attenuation law for both, following \citet{Lyu2023}, we use two different dust laws, namely, the dust attenuation law of \citet{Calzetti2000}  for the host galaxy and the dust extinction curve of SMC \citep{Gordon2003} for the AGN. Finally, instead of assuming a nonparametric SFH, we use a parametric delayed-tau SFH. 

\begin{figure}
    \centering
    \includegraphics[width=0.7\textwidth]{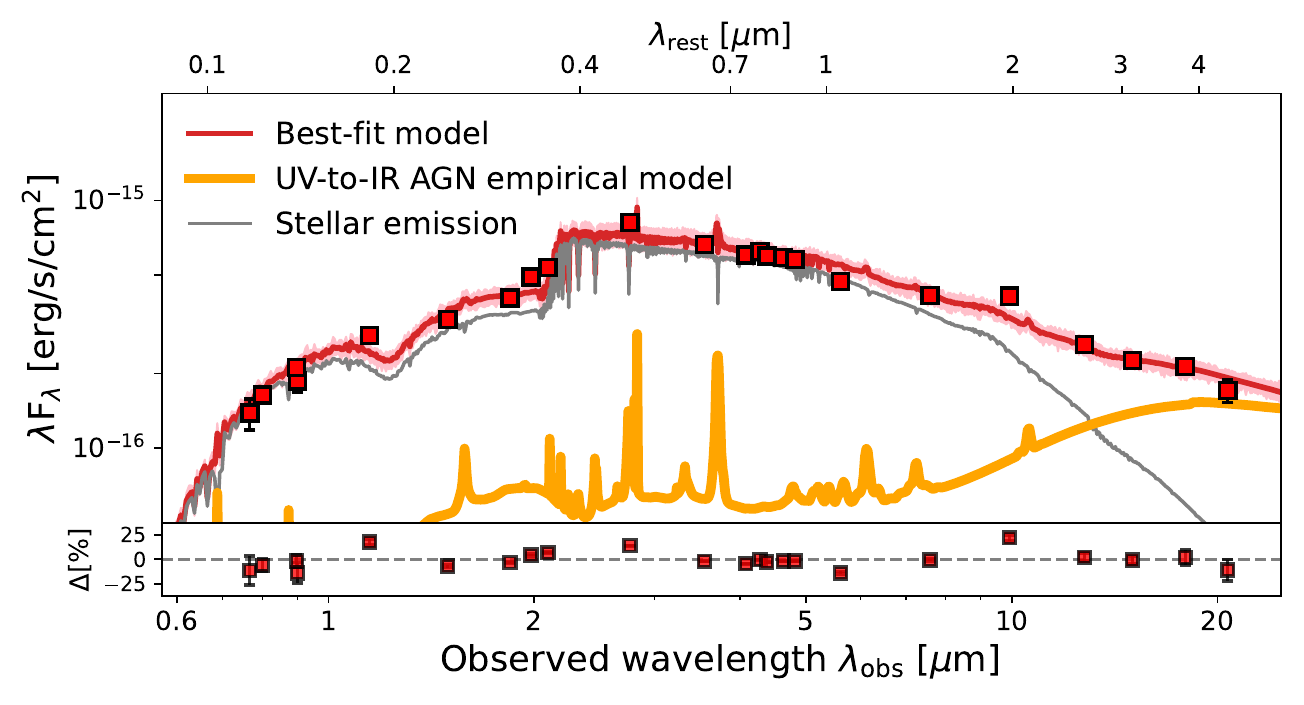}
    \caption{Alternative SED fitting of the inner region of GS-9209. Despite that the SED model assumptions here differ greatly from the fiducial ones used in the main text (see Section \ref{method:sed_inner} for details), our key conclusions stand: For the inner region of GS-9209, the observed MIRI fluxes (1) require the addition of hot dust emission to reproduce and (2) are dominated by the dust emission of AGN.}
    \label{fig:alter_sed_inner}
\end{figure}

Fig. \ref{fig:alter_sed_inner} shows the fitting results of the alternative SED modeling. To begin, unlike the \citet{Nenkova2008a, Nenkova2008b} model which only includes the contribution from AGN torus, the semi-empirical templates of \citet{Lyu2023} actually also include the UV-to-optical AGN emission. Yet, the fitting still suggests that the rest-UV to NIR emission of GS-9209 is dominated by starlight. At \lamrest $\gtrsim2\mu m$, similar to what we found in the fiducial modeling, the observed MIRI fluxes cannot be reproduced by the alternative model with stellar emission alone, i.e. the need for dust emission. We remind that the dust emission of diffuse ISM is included to the alternative SED modeling. Nonetheless, the fitting still suggests that GS-9209's SED at \lamrest $\gtrsim3\mu m$ is dominated by the dust emission of AGN torus, in line with the fiducial modeling. It is also worth mentioning that based on our SED fitting we estimate the AGN luminosity of GS-9209 at rest-frame 5100\AA\ to be $L_{5100}=10^{10.17}L_\odot$ which is in great agreement with that independently estimated by \citet{Carnall2023} who obtained $L_{5100}\sim10^{10}L_\odot$ by SED fitting using the NIRSpec spectrum or $L_{5100}\sim10^{10.2}L_\odot$ by converting the observed broad H$\alpha$ flux using the relation of \citet{Greene2005}.

Finally, another line of evidence that the inner MIRI fluxes of GS-9209 at \lamrest $\gtrsim3\mu m$ is dominated by AGN comes from additional SED fitting where we switch off the AGN component. Without the AGN component, the best-fit SED has the total (i.e. summation of all bands) $\chi^2_{tot}=56.3$, which, as expected, is worse compared to $\chi^2_{tot}=37.4$ of the best-fit model with AGN switched on. More importantly, with the assumption of energy balance, i.e. the energy absorbed by dust in the rest-frame UV is re-emitted in the IR, such a fitting forces to use the dust emission associated with diffuse ISM (i.e. star formation) to explain the observed MIRI fluxes. As a result, this fitting returns SFR $=50\pm5M_\odot$/yr, which is significantly higher than both that inferred from the narrow H$\alpha$ emission \citep{Carnall2023} (SFR$_{\rm{H\alpha, narrow}}\sim2M_\odot$/yr), and the strict upper limit set by archival ALMA non-detections of GS-9209 \citep{Santini2021} (SFR$_{\rm{ALMA}}<41M_\odot$/yr). These together strongly argue against the origin of the observed MIRI fluxes to be dominated by the dust emission associated with diffuse ISM.

To summarize, the assumptions made in the above alternative SED modelings differ greatly from the fiducial ones, which we did intentionally. Yet, they reach to the conclusions fully in line with the fiducial SED fitting. Therefore, we conclude that our results regarding the origin of MIRI fluxes of GS-9209's inner region -- AGN dominated -- are not sensitive to the SED assumptions.

\subsection{SED modeling of the outer region}

Here we describe our SED analysis of the outer region of GS-9209. The outer region is enclosed by a $0"3<r<0.7"$ circular annulus aperture. Due to PSF broadening, we subtract the flux contribution from the inner part to the outer aperture by assuming the inner component is a point source, which is a good approximation because the surface brightness profile of GS-9209's inner part is very similar to PSF (Fig. \ref{fig:sb_miri}). As the flux contribution from the inner region to the outer aperture has been removed, the AGN component is not included to the SED fitting of the outer region.

\begin{figure}
    \centering
    \includegraphics[width=0.7\textwidth]{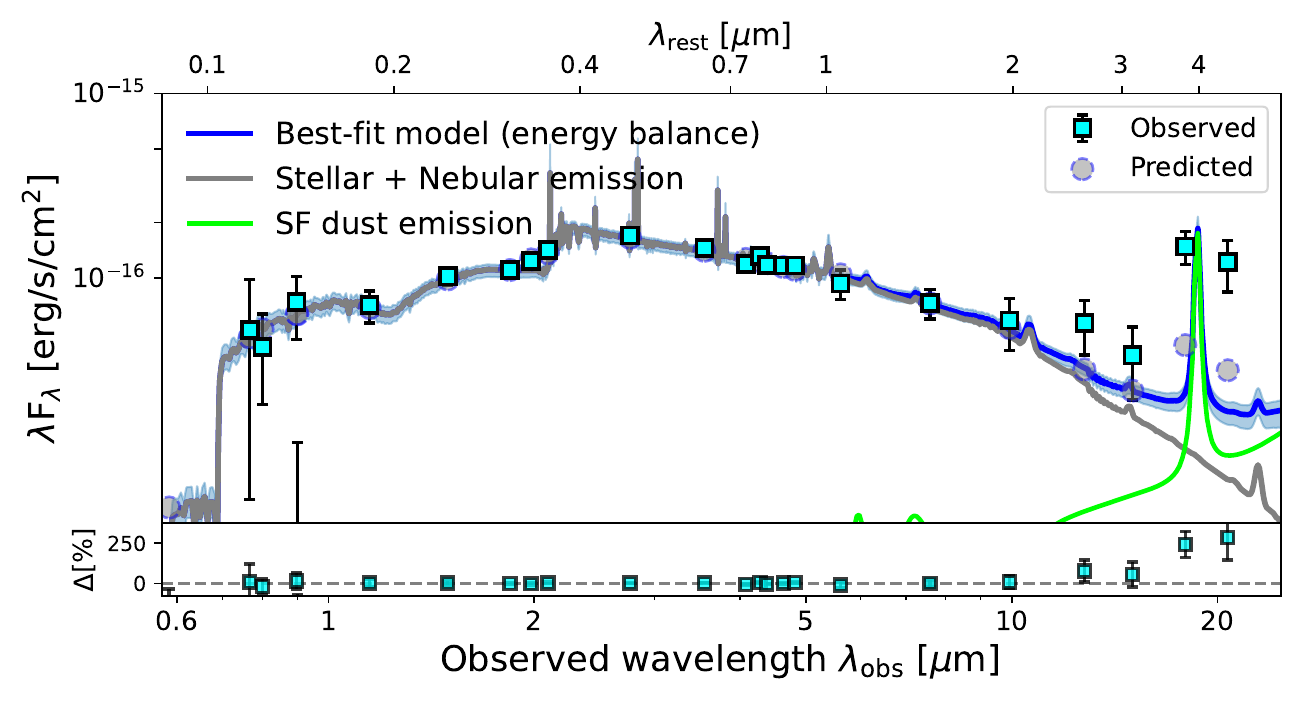}
    \caption{SED fitting of the outer region of GS-9209 with the assumption of energy balance. The observed MIRI fluxes at \lamobs $>10\mu m$ are not well fitted under this assumption.}
    \label{fig:sed_outer_eb}
\end{figure}

To begin, we run SED fitting with the assumption of energy balance. By default, we use the \citet{Draine2007} model. As Fig. \ref{fig:sed_outer_eb} shows, the observed MIRI fluxes at \lamobs $>10\mu m$, or \lamrest $>2.5\mu m$, cannot be fitted well. This issue cannot be mitigated by replacing the \citet{Draine2007} model with other (semi-)empirical dust models extensively used in the literature, e.g. \citep{Rieke2009,Lyu2016}. Further, we test the finding by using a different stellar synthesis code {\sc BC03} \citep{Bruzual2003} and SED fitting software {\sc Bagpipes} \citep{Carnall2018}. These changes do not help to mitigate the issue, either. As detailed in the main text, however, this issue can be largely mitigated, if not fully resolved, by removing the energy balance assumption. 

\subsection{Tests on other SED assumptions}

\begin{figure}
    \centering
    \includegraphics[width=0.97\textwidth]{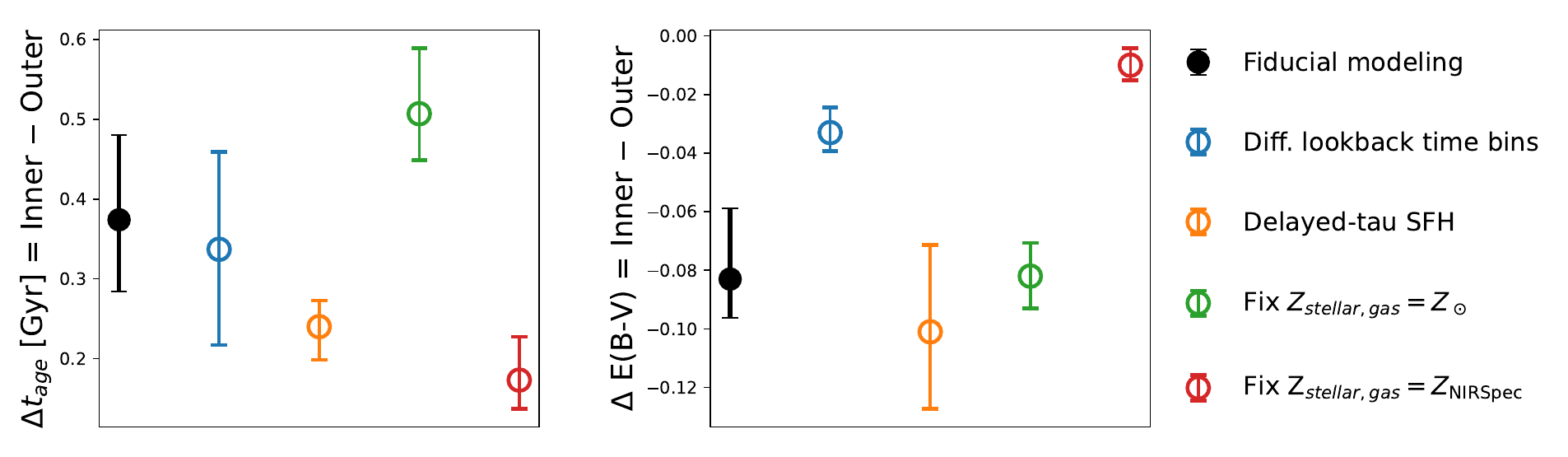}
    \caption{The differences in stellar age and $\rm{E(B-V)}$ between GS-9209's inner and outer regions from SED fittings with SFH and metallicity assumptions differing from the fiducial ones. Our conclusions do not change: The outer region of GS-9209 is younger ($\Delta t_{\rm{age}}>0$) and more dust attenuated ($\Delta\rm{E(B-V)}<0$) than its inner region. }
    \label{fig:sed_other_assumption}
\end{figure}

We now investigate the impact of other SED assumptions on the key conclusions of our SED analysis, namely that the outer region of GS-9209 is younger and more dust attenuated than its inner region. The results are plotted in Fig. \ref{fig:sed_other_assumption}. Specifically, we test the assumptions of:
\begin{itemize}
	\item SFH. We test our results by using two other different forms of SFH, namely (1) a parametric delayed-tau SFH and (2) a nonparametric continuity SFH with lookback time bins different from the fiducial ones (here the nine lookback time bins are assumed to be $0-30$ and $30-100$ Myr for the first two bins; $0.85 t_{\rm{H}} - t_{\rm{H}}$ for the last bin; and evenly spaced in the logarithmic lookback time between 100 Myr and $0.85 t_{\rm{H}}$ for  the remaining six bins). As Fig. \ref{fig:sed_other_assumption} (blue and orange circles) shows, we do not see any substantial changes in our conclusions by altering the assumed SFH.
	\item metallicities. Instead of setting the stellar and gas-phase metallicities of GS-9209 to be free parameters, we assume that the inner and outer regions have the same metallicities and we fix them to be either (1) the best-fit values measured with NIRSpec spectroscopy \citep{Carnall2023} or (2) the solar values. We note that, according to its NIRSpec spectrum, the stellar metallicity of GS-9209 should be substantially lower than the solar value \citep{Carnall2023} and there is no evidence that the metallicities of the inner and outer regions are the same. As Fig. \ref{fig:sed_other_assumption} (green and red circles) shows, even with these arguably ``bad'' metallicity assumptions, we still find that the outer region of GS-9209 is younger and more dust attenuated than its inner region, showing that our conclusions are not sensitive to the assumed metallicities.
\end{itemize}

\subsubsection*{Acknowledgments}

ZJ, GHR, FS, JMH, MR, YZ and CNAW are supported by JWST/NIRCam contract to the University of Arizona NAS5-02015.
The research of CCW is supported by NOIRLab, which is managed by the Association of Universities for Research in Astronomy (AURA) under a cooperative agreement with the National Science Foundation.
GHR, JL and SA acknowledge support from the JWST Mid-Infrared Instrument (MIRI) Science Team Lead, grant 80NSSC18K0555, from NASA Goddard Space Flight Center to the University of Arizona.
FD and RM acknowledge support by the Science and Technology Facilities Council (STFC), ERC Advanced Grant 695671 ``QUENCH", and by the UKRI Frontier Research grant RISEandFALL. RM also acknowledges funding from a research professorship from the Royal Society.
ST acknowledges support by the Royal Society Research Grant G125142.
BER acknowledges support from the NIRCam Science Team contract to the University of Arizona, NAS5-02015, and JWST Program 3215.
AJB acknowledges funding from the ``FirstGalaxies" Advanced Grant from the European Research Council (ERC) under the European Union’s Horizon 2020 research and innovation programme (Grant agreement No. 789056).
The authors acknowledge the FRESCO team led by Dr. Oesch for developing their observing program with a zero-exclusive-access period.
This work made use of the lux supercomputer at UC Santa Cruz which is funded by NSF MRI grant AST 1828315, as well as the High Performance Computing (HPC) resources at the University of Arizona which is
funded by the Office of Research Discovery and Innovation (ORDI), Chief Information Officer (CIO), and University Information Technology Services (UITS).

\subsubsection*{Author contributions}
ZJ, CCW, GHR, JL and SA contributed to the initial discovery. All authors contributed to the interpretation of results. 
ZJ, JL and JMH contributed to the analysis of Spectral Energy Distribution. ZJ, CCW and FS contributed to the morphological analysis. 
GHR, JL, SA, IS, YZ contributed to the design and data reduction of the MIRI imaging of the SMILES program. MR, FS, FD, ST, BR, RM, AJB and CNAW contributed to the design and data reduction of the NIRCam imaging of the JADES program. CCW and ST contributed to the design and data reduction of the NIRCam imaging of the JEMS program.

\bibliography{sn-bibliography}

\end{document}